\documentclass[conference, 9pt]{IEEEtran}

\newif\iftr     
\trtrue

\newif\ifconf   
\conffalse

\newif\ifblind  
\blindfalse

\newif\ifcom    
\comfalse 

\newif\ifps     
\psfalse

\usepackage{cite}
\usepackage{amsmath,amssymb,amsfonts}
\usepackage{algorithmic}
\usepackage{graphicx}
\usepackage{textcomp}
\usepackage{xcolor}

\usepackage{subcaption}     
\usepackage{makecell}       
\usepackage{multirow}       
\usepackage{color, colortbl}
\usepackage{pifont}         
\usepackage{enumitem}		
\usepackage{booktabs}       
\usepackage{cleveref}       
\usepackage{glossaries}     
\usepackage{microtype}
\usepackage{siunitx}
\usepackage{tikz}
\usepackage{stfloats}

\ifcom
\usepackage{color-edits}    		
\addauthor[Summary]{ps}{orange}
\addauthor[Patrick]{pi}{teal}
\addauthor[Maciej]{mb}{blue}
\addauthor[Torsten]{th}{red}
\addauthor[Matheus]{mc}{orange}
\addauthor[Tim]{tf}{violet}
\fi

\setacronymstyle{long-short}
\newacronym{rtl}{RTL}{register transfer level}
\newacronym{noc}{NoC}{network-on-chip}
\newacronym{nocs}{NoCs}{networks-on-chip}

\def\BibTeX{{\rm B\kern-.05em{\sc i\kern-.025em b}\kern-.08em
    T\kern-.1667em\lower.7ex\hbox{E}\kern-.125emX}}

\definecolor{lightgray}{gray}{0.8}
\newcommand{\undergrayed}[1]{\tikz[baseline=(X.base)]\node [draw=none,fill=lightgray,text=black,semithick,rectangle,inner sep=1pt, rounded corners=3pt] (X) {#1};}

\def \nprinciples {four }
    
\newcommand{\picom}[1]{\ifcom\picomment{#1}\fi}

\newcommand{\ps}[1]{\ifps\pscomment{#1}\fi}

\setlist{leftmargin=1em}
\newcommand{\cmark}{\ding{52}}  
\newcommand{\xmark}{\ding{56}}  
\newcommand{\pone}{\ding{182}}   
\newcommand{\ptwo}{\ding{183}\hspace{0.1em}}   
\newcommand{\pthree}{\ding{184}\hspace{0.1em}} 
\newcommand{\pfour}{\ding{185}\hspace{0.1em}}  

\begin{document}

\title{Sparse Hamming Graph:\\A Customizable Network-on-Chip Topology\ifconf\vspace{-2em}\fi}

\ifblind
\author{}
\else
\author{
\IEEEauthorblockN{
Patrick Iff\IEEEauthorrefmark{1}, 
Maciej Besta\IEEEauthorrefmark{1}, 
Matheus Cavalcante\IEEEauthorrefmark{2},
Tim Fischer\IEEEauthorrefmark{2}, 
Luca Benini\IEEEauthorrefmark{2}\IEEEauthorrefmark{3} and
Torsten Hoefler\IEEEauthorrefmark{1}
}
\IEEEauthorblockA{
\IEEEauthorrefmark{1}Department of Computer Science,
ETH Zurich, Zurich, Switzerland\\
Email: \{patrick.iff, maciej.besta, htor\}@inf.ethz.ch}
\IEEEauthorblockA{
\IEEEauthorrefmark{2}Department of Information Technology and Electrical Engineering,
ETH Zurich, Zurich, Switzerland\\
Email: \{matheus, fischeti, lbenini\}@iis.ee.ethz.ch}
\IEEEauthorblockA{
\IEEEauthorrefmark{3}
Dept. of Electrical, Electronic and Information Engineering, University of Bologna, Italy
\iftr
\vspace{-1em}
\fi}
}
\fi

\maketitle
\begin{abstract}
Chips with hundreds to thousands of cores require scalable networks-on-chip (NoCs).
Customization of the NoC topology is necessary to reach the diverse design goals of different chips.
We introduce sparse Hamming graph, a novel NoC topology with an adjustable cost-performance trade-off
that is based on \nprinciples NoC topology design principles we identified.
To efficiently customize this topology, we develop a toolchain that leverages 
approximate floorplanning and link routing to deliver fast and accurate cost 
and performance predictions.
We demonstrate how to use our methodology to achieve desired cost-performance trade-offs
while outperforming established topologies in cost, performance, or both.
\end{abstract}
%
%
\section{Introduction}
\label{sec:intro}

\ps{We need high-performance NoCs}

CMOS technology scaling keeps increasing the transistor density, 
enabling the integration of a growing number of compute cores into modern processors and accelerators.
While the parallel performance of such architectures is skyrocketing, more and more applications are becoming bound by data movement
\iftr
\cite{datamovement, lumsdaine2007challenges, besta2020communication, besta2017push, sakr2020future, besta2019slim}.
\else
\cite{datamovement, besta2020communication}.
\fi
\iftr
This includes traditional graph processing kernels \cite{besta2017slimsell, besta2020high, gianinazzi2018communication},
dense linear algebra kernels \cite{kwasniewski2019red, kwasniewski2021parallel, kwasniewski2021pebbles}, 
complex pattern matching applications~\cite{besta2021sisa, besta2021graphminesuite, besta2022probgraph, chakrabarti2012graph, rehman2012graph, ramraj2015frequent, cliques, besta2021motif},
but also newer classes of mixed sparse-dense workloads such as graph neural networks \cite{besta2022parallel, kipf2016semi, besta2022neural, zhou2020graph, wu2020comprehensive}.
\fi
\iftr
These developments stress
\else
This stresses
\fi
the importance of scalable, high-performance \gls{nocs}. 

\ps{Introduce design principles}

A key decision to make when designing a \gls{noc} is the choice of a suitable topology of links between routers. 
The NoC topology has a large influence on the NoC's performance (latency, throughput) and cost (area, power).
To guide the design of low-cost and high-performance \gls{noc} topologies, we identify \nprinciples fundamental \gls{noc} topology design principles 
(Section~\ref{sec:insights}, \textbf{contribution \#1}).
To achieve low cost, \pone~\emph{use low-radix topologies} and \ptwo~\emph{design for link routability} and 
to achieve high performance, \pthree~\emph{minimize the network diameter} and 
\pfour~\emph{minimize the physical path length}\footnote{The total physical distance that the data travels across the chip.}.
Principle \pone~minimizes the router area and the overall number of links and 
principle \ptwo~minimizes the area overhead induced by links - both minimizing the NoC's cost.
Principle \pthree~minimizes the number of router-to-router hops per flit (minimize latency) 
and hence the number of flits processed by a given router (minimize congestion $\to$ maximize throughput).
Principle \pfour~minimizes the latency which is linear in the physical path length.

\ps{Motivate and introduce sparse Hamming graph}

Achieving a low network diameter (following \pthree) requires a high router radix (violating \pone) and vice versa.
Hence, no single topology can provide high-performance and low-cost, instead, a trade-off between the two is needed.
To meet the diverse requirements of different architectures, the NoC topology's cost-performance trade-off must be adjustable.
We introduce \textbf{sparse Hamming graph}, a customizable NoC topology
that is based on our \gls{noc} topology design principles (Section~\ref{sec:class}, \textbf{contribution \#2}). 
By adjusting the sparsity of links in the topology, we can steer its cost-performance trade-off.

\ps{Motivate and introduce toolchain}

Different architectures do not only differ in their design goals (e.g., low-power vs. high-performance) 
but also in their architectural parameters (e.g., number and size of cores, technology node, or transport protocol).
We want to adjust the sparsity of the sparse Hamming graph topology to both, the design goals and the architectural parameters.
To do this, we need a way to predict the performance and cost that a NoC topology attains if applied to a given architecture.
Existing models for cost and performance predictions can be grouped into high-level and low-level models.
High-level models are fast, but they do not consider implementation details like, e.g., link routing, which results in a lack of accuracy.
Low-level models are accurate but extremely time-consuming which renders them unwieldy for a quick exploration of a large design space.
To bridge the gap between high-level and low-level models, 
we develop a novel, custom cost and performance prediction toolchain\footnote{Source code: https://spclgitlab.ethz.ch/iffp1/sparse\_hamming\_graph\_public} 
(Section~\ref{sec:model}, \textbf{contribution \#3}).
Our toolchain works at the speed of high-level models but other than existing high-level models, 
it estimates implementation-details by performing approximate floorplanning and link routing.

\ps{Summarize evaluatoin results: Customizable + Pareto-Optimal}

To evaluate the proposed NoC topology customization strategy, we use our prediction toolchain to rapidly customize 
four sparse Hamming graph topologies for four different architectures, each with its unique set of architectural parameters.
The design goal is to maximize each \gls{noc}'s performance without dedicating more than $40\%$ of the total chip area to it.
Our evaluation (Section \ref{sec:evaluation}) shows that 
our customized sparse Hamming graph topologies outperform established topologies in terms of cost, performance, or both.

\section{Design Principles for \gls{noc} Topologies}
\label{sec:insights}

\begin{figure*}[t]
\centering
\captionsetup{justification=centering}
\begin{subfigure}[t]{0.130 \textwidth}
\centering
\includegraphics[width=1.05\columnwidth]{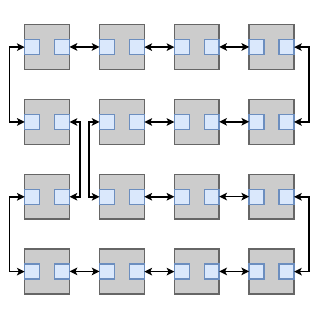}
\caption{Ring:\\Links form a cycle.}
\label{fig:insights-topo-ring}
\end{subfigure}
$\mkern3mu$
\begin{subfigure}[t]{0.130 \textwidth}
\centering
\includegraphics[width=1.05\columnwidth]{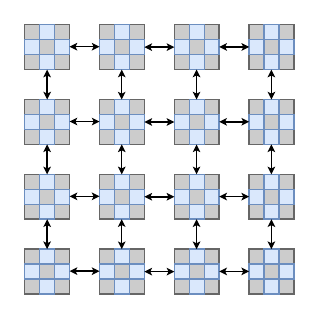}
\caption{2D Mesh:\\Neighboring tiles are connected.}
\label{fig:insights-topo-mesh}
\end{subfigure}
$\mkern3mu$
\begin{subfigure}[t]{0.130 \textwidth}
\centering
\includegraphics[width=1.05\columnwidth]{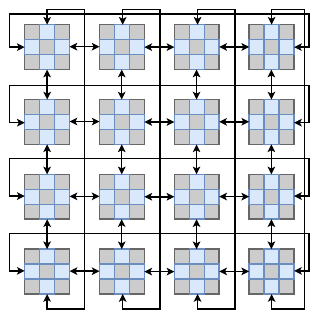}
\caption{2D Torus:\\Rows/columns form a cycle.}
\label{fig:insights-topo-torus}
\end{subfigure}
$\mkern3mu$
\begin{subfigure}[t]{0.130 \textwidth}
\centering
\includegraphics[width=1.05\columnwidth]{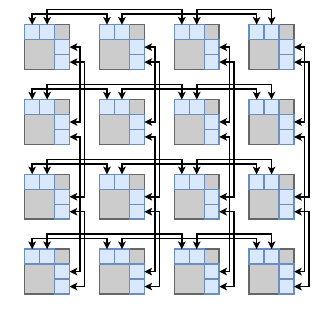}
\caption{Fld. 2D Torus:\\2D torus without long links.}
\label{fig:insights-topo-foldedtorus}
\end{subfigure}
$\mkern3mu$
\begin{subfigure}[t]{0.130 \textwidth}
\centering
\includegraphics[width=1.05\columnwidth]{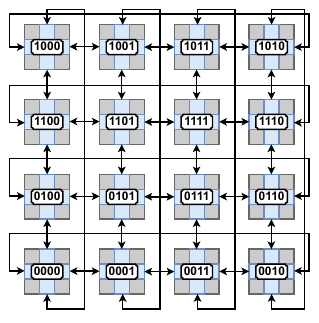}
\caption{Hypercube:\\Connect tiles if IDs differ in 1 bit.}
\label{fig:insights-topo-hypercube}
\end{subfigure}
$\mkern3mu$
\begin{subfigure}[t]{0.130 \textwidth}
\centering
\includegraphics[width=1.05\columnwidth]{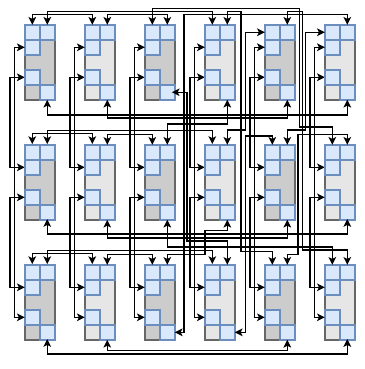}
\caption{SlimNoC \cite{slimnoc}:\\Topology based on MMS graphs.}
\label{fig:insights-topo-slimnoc}
\end{subfigure}
\begin{subfigure}[t]{0.130 \textwidth}
\centering
\includegraphics[width=1.05\columnwidth]{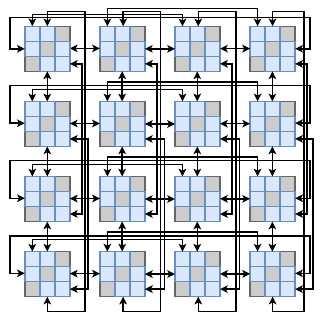}
\caption{Flat. Butterfly:\\Fully-connected rows/columns.}
\label{fig:insights-topo-flattenedbutterfly}
\end{subfigure}
\caption{Visualization of established \gls{noc} topologies. Gray squares represent tiles, blue squares represent 
ports, and arrows represent links.}
\label{fig:topologies}
\vspace{-1.5em}
\end{figure*}

\ps{2 design principles to reduce cost}

\ps{Assumptions: Direct topology + Don't route over tiles}

\subsection{Assumptions}
\label{ssec:insights-assumptions}

Throughout the whole paper, we assume a chip that is organized as an $R \times C$ grid of identical building blocks which we call tiles.
Each tile contains one or more endpoints (e.g., compute cores or memories) and a local router to which all of its endpoints are connected. 
A \gls{noc} is used to provide connectivity between different tiles.
The \gls{noc}'s links are attached to the tile's local routers.
This regular tiled design is common for large compute fabrics as its modularity keeps the physical design effort at the top level under control.
We assume that each metal layer has a predefined routing direction.
Furthermore, whenever a link is too long to be operated at the target clock frequency, we insert as many registers (pipeline stages) as necessary to meet said frequency. 
As a result, sending a flit across a link can take multiple clock cycles \cite{noc-multicycle-links} which means that we need network components \cite{xpipes, axinoc} capable of handling multi-cycle links.
Finally, we assume that tiles occupy all available metal layers which disallows routing an inter-tile link between two tiles $A$ and $B$ over a third tile $C$.

\subsection{Reduce Cost (Area, Power)}
\label{ssec:insights-area}

The following two design principles aim at reducing the \gls{noc}'s area. 
Since a chip's power consumption is roughly linear in its area \cite{vlsi-textbook}, 
following these design principles also reduces the power consumption.

\undergrayed{\textbf{\pone~Use Low-Radix Topologies:}}
The area of most router architectures scales quadratically with the router 
radix \cite{dpin}\picom{Citation suggested by TH - The book doesn't seem to be downloadable for free}, 
hence, high-radix topologies come with large area requirements.
Additionally, in a setting with one router per tile, a higher router radix 
implies a higher overall number of links which further inflates the \gls{noc}'s area.

\undergrayed{\textbf{\ptwo~Design for Routability:}}
When designing a \gls{noc} topology, it is crucial to pay attention to the routability of links 
which can be improved by meeting the following four criteria:
\begin{itemize}
	\item	\textbf{Short Links}: 
			Links between adjacent tiles come with minuscule area 
			overheads while long links occupy a large area.
	\item	\textbf{Aligned Links}:
			Links between tiles in the same row or column tend to be more area-efficient 
			than links that change row and column.
	\item	\textbf{Uniform Link Density}:
			Assume that the space between two rows or columns of tiles contains sections with 
			high and sections with low link density (like, e.g., SlimNoC, see 
			Figure \ref{fig:insights-topo-slimnoc}). The spacing between said rows or
			columns needs to be large enough to fit all links from the high-density 
			section, resulting in an under-utilization of the low-density section. 
			To use the area efficiently, we should aim for a uniform link density
			(like, e.g., 2D torus, see Figure \ref{fig:insights-topo-torus}).
	\item	\textbf{Optimized Port Placement}: The location of ports within a tile influences 
			the routability of links. E.g., for a mesh topology, we can place one
			port on each face of the tile which results in short, straight links between 
			tiles (see Figure \ref{fig:insights-topo-mesh}). In the ring topology on the other
			hand, the ports are placed north/south or east/west which both 
			inflates the lengths of a subset of links (see Figure \ref{fig:insights-topo-ring}).
\end{itemize}

\ps{2 design principles to increase performance}

\subsection{Boost Performance (Latency, Throughput)}
\label{ssec:insights-latency}

\undergrayed{\textbf{\pthree~Minimize the Network Diameter:}}
The network diameter is the maximum number of routers-to-router hops that a flit\footnote{Flow control unit: Atomic amount of data transported across the network.}
takes on the path from its source-tile to its destination-tile. Whenever a flit traverses a router, 
it experiences a certain delay, therefore, minimizing the network diameter reduces the latency.
Furthermore, a reduction of the average number of router-to-router hops per flit goes along with a reduction of the average
number of flits processed by a given router which reduces the congestion at routers and improves the throughput.

\undergrayed{\textbf{\pfour~Minimize the Physical Path Length:}}
The delay of a link built of buffered wires is linear in its length \cite{vlsi-textbook}.
As a consequence, minimizing the physical distance that a flit travels is crucial to achieve low latency.
The prerequisite for using paths of minimal physical length is the existence of links that provide such paths.
However, the pure existence of these links is not enough since the routing algorithm might send flits over physically non-shortest paths.
This could for example happen if the routing algorithm minimizes the number of router-to-router hops, which does not necessarily minimize the physical path length.
Adapting the routing algorithm to minimize the physical path length is not the solution since, e.g., in a 2D torus 
topology (see Figure \ref{fig:insights-topo-torus}), this would leave the wrap-around links unused and severely deteriorate the throughput.
To provide low latency while sustaining high throughput, a \gls{noc} topology should 
a) contain links that provide physically minimal paths and 
b) be co-designed with the routing algorithm to use these paths without reducing the throughput.

\section{The Sparse Hamming Graph \gls{noc} Topology}
\label{sec:class}

\ifconf
\begin{table*}[bp]
\else
\begin{table*}[h]
\fi
\vspace{-0.5em}
\setlength{\tabcolsep}{2.5pt}
\centering
\captionsetup{justification=centering}
\caption{
Compliance of \gls{noc} topologies for $R$ rows and $C$ columns of tiles with our design principles.
For sparse Hamming graph, we provide intervals for achievable values and
we write (\cmark) if a design principle is only achieved for some parametrizations.
\textbf{SL}: Short Links, 
\textbf{AL}: Aligned Links, 
\textbf{ULD}: Uniform Link Density, 
\textbf{OPP}: Optimized Port Placement.
$^{*}$Paths with minimal physical length, 
$^{**}$The links at hand provide minimal paths, 
$^{***}$Routing algorithms that minimize the number of router-to-router hops use  minimal paths.
$^{\dagger}$Hypercube is only applicable if $R$ and $C$ are powers of two.
$^{\ddagger}$SlimNoC is only applicable if $RC = 2p^2$ for a prime power $p$.
}
\begin{tabular}{lccccccccc}

\hline
\multirow{2}{*}[-0.2em]{Topology} & 
\multirow{2}{*}[-0.2em]{\makecell{Router \\ Radix}} & 
\multicolumn{4}{c}{Design for Routability} &
\multirow{2}{*}[-0.2em]{\makecell{Network \\ Diameter}} &
\multicolumn{2}{c}{Minimal Paths$^{*}$} &
\multirow{2}{*}[-0.2em]{\makecell{\#Configurations\\for given $R$ and $C$}} \\
& & SL & AL & ULD & OPP & & Present$^{**}$ & Used$^{***}$\\

\midrule
Ring 				
& 2							& \cmark	& \cmark 	&$\thicksim$&	\xmark	& $RC/2$		& \xmark	& \xmark & 1\\
2D Mesh 				
& 4							& \cmark	& \cmark 	& \cmark	&	\cmark	& $R+C-2$		& \cmark	& \cmark & 1\\
2D Torus			
& 4							& \xmark	& \cmark 	& \cmark	&	\cmark	& $R/2+C/2$		& \cmark	& \xmark & 1\\
Folded 2D Torus	\cite{folded-torus}
& 4							&$\thicksim$& \cmark 	& \cmark	&	\cmark	& $R/2+C/2$		& \xmark	& \xmark & 1\\
Hypercube \cite{hypercube}
& $log_2{(RC)}$ 			& \xmark	& \cmark	& \cmark	& 	\cmark	& $log_2{(RC)}$	& \cmark	& \xmark & 0 or 1$^{\dagger}$\\
SlimNoC \cite{slimnoc}
& $\approx \sqrt{RC}$		& \xmark	& \xmark	& \xmark	& 	\xmark	& $2$			& \xmark	& \xmark & 0 or 1$^{\ddagger}$\\
Flattened Butterfly
\cite{flattened-butterfly-noc}
& $R+C-2$ 					& \xmark	& \cmark	& \xmark	& 	\cmark	& $2$			& \cmark	& \cmark & 1\\
\midrule
\makecell[l]{Sparse Hamming Graph (This Work)} & 
$[4,R+C-2]$  &
(\cmark) &
\cmark &
(\cmark) &
\cmark &
$[2,R+C-2]$ &
\cmark &
(\cmark) &
$2^{R+C-4}$ \\
\hline
\end{tabular}
\label{tab:topologie-properties}
\end{table*}

\ps{Section intro}

We propose the customizable sparse Hamming graph topology that is based on our \nprinciples \gls{noc} topology design principles.
The sparsity of this topology depends on two input parameters which enables an effortless adjustment of its cost-performance trade-off.
Table \ref{tab:topologie-properties} shows the compliance of sparse Hamming graph and many 
established \gls{noc} topologies with our \nprinciples \gls{noc} topology design principles.

\ps{Concept: Span design space Mesh - Flattened Butterfly}

\paragraph{Concept}
The base for a sparse Hamming graph is a 2D mesh topology. It fulfills all design for routability
criteria (\ptwo) and it provides a low router radix (\pone) which ensures a low cost.
In addition, the mesh topology provides paths with minimal physical length (\pfour) resulting in
low latency. The downside of the mesh topology is the relatively high network diameter (\pthree)
that leads to significant performance penalties when scaling up the number of tiles. 
To alleviate this issue, we add additional links to our topology. 
The number and shape of these additional links are configured through a set of
parameters. The way in which we add these links (refer to the next paragraph 
for details) ensures that a good routability of links (\ptwo) is maintained. 
With the maximum number of additional links, we reach a flattened butterfly topology 
that provides a low network diameter (\pthree) and excellent performance. 
The downside of the flattened butterfly is its significant area overhead.
In summary, the sparse Hamming graph spans the design space
between a mesh topology (low cost) and a flattened butterfly topology (high performance) while
following our design principles to ensure reaching favorable cost-performance trade-offs.

\begin{figure}[h]
\vspace{-1em}
\centering
\captionsetup{justification=centering}
\includegraphics[width=0.8\linewidth]{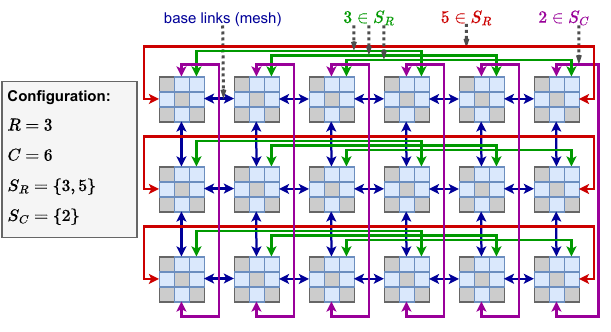}
\caption{Construction scheme and example for sparse Hamming graph.}
\label{fig:class-of-topologies}
\vspace{-1em}
\end{figure}

\ps{Construction: Mesh + parameterized skip-links}

\paragraph{Construction}
Let $R$ and $C$ be the number of rows and columns of tiles in our chip.
The sparse Hamming graph topology takes two sets as parameters: A set 
$S_{R} \subseteq \{x \in \mathbb{N}~|~2 \leq x < C\}$ and a set 
$S_{C} \subseteq \{x \in \mathbb{N}~|~2 \leq x < R\}$.
Let $T_{r,c}$ be the tile in row $r$ and column $c$. 
We start with a 2D mesh topology. For each row $r$, for each number 
$x \in S_{R}$ and for each number $i \in \mathbb{N}$, $1 \leq i \leq (C-x)$,
we add a link $T_{r,i} \leftrightarrow T_{r,i+x}$.
Likewise, for each column $c$, for each number $x \in S_{C}$ and for each number 
$i \in \mathbb{N}$, $1 \leq i \leq (R-x)$, we add a link $T_{i,c} \leftrightarrow T_{i+x,c}$. 
Figure \ref{fig:class-of-topologies} visualizes this construction.
Since all topologies that are constructed this way are subgraphs of a 
2D Hamming graph, we call this topology \textit{sparse Hamming graph}.
\picom{Do we need to site 2D Hamming graphs? I didn't find a good citation.}
\section{Prediction of Performance and Cost}
\label{sec:model}

\ps{Section intro}

To efficiently customize the sparse Hamming graph topology to the design goals and architectural parameters at hand,
we develop our custom toolchain for performance and cost predictions.

\subsection{Overview of Prediction Toolchain}
\label{ssec:model-toolchain}

\ps{Toolchian overview: Model + BookSim}

\Cref{fig:model-toolchain} shows our toolchain for \gls{noc} performance
and cost predictions. We develop a custom model that leverages 
a set of architectural parameters to estimate the area and power consumption (cost)
associated with a given \gls{noc} topology. In addition, our model estimates the latency of
every router-to-router link in the \gls{noc}. We feed the \gls{noc} topology with link latency
estimates as well as information about the router architecture,
routing algorithm, and traffic pattern into the cycle-accurate BookSim2 
\cite{booksim} \gls{noc} simulator. Based on the results of the BookSim2 simulations, we estimate the \gls{noc}'s
zero-load latency and saturation throughput (performance). The key to accurate simulations in
BookSim2 are our model's link latency estimates.
\begin{figure}[h]
\centering
\captionsetup{justification=centering}
\includegraphics[width=1.0\linewidth]{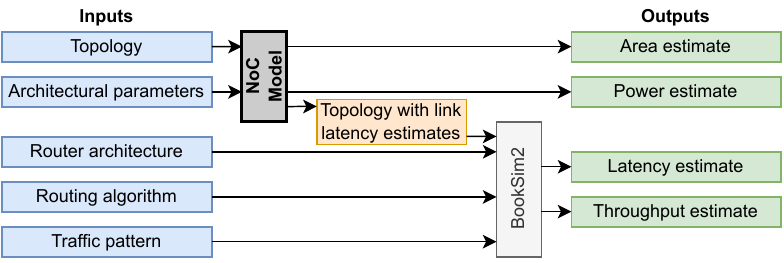}
\caption{Toolchain to predict performance and cost metrics of a \gls{noc}.}
\label{fig:model-toolchain}
\vspace{-0.5em}
\end{figure}

\subsection{A Model for Area, Power and Link Latency Prediction}
\label{ssec:model-custom}

\ps{Model overview}

\Cref{fig:model} visualizes our model for predicting a \gls{noc}'s area overhead, power 
consumption, and the latency of every router-to-router link. Our model is based on 
the assumptions stated in Section \ref{ssec:insights-assumptions}.
\begin{figure}[h]
\centering
\captionsetup{justification=centering}
\includegraphics[width=1.0\linewidth]{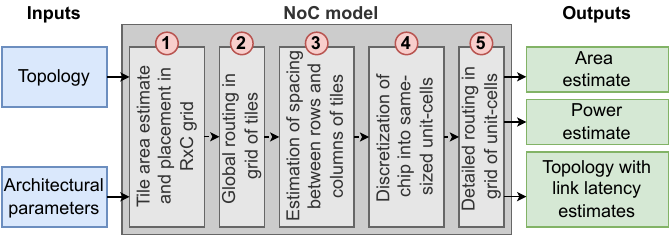}
\caption{Our model to predict area overhead, power consumption, and link latencies of 
a \gls{noc}.}
\label{fig:model}
\vspace{-0.5em}
\end{figure}

\ps{Model inputs}

\subsubsection{Architectural Parameters}
\label{sssec:model-custom-archparam}

\Cref{tab:model-details-archparam} summarizes the architectural parameters that our 
model takes as inputs. The descriptions in \Cref{tab:model-details-archparam} make
it straightforward to set most of these parameters. Configuring the functions 
$f^H_\text{wires $\to$ mm}(x)$ and $f^V_\text{wires $\to$ mm}(x)$ that return the 
channel width (in mm) needed to build $x$ horizontal and $x$ vertical wires respectively is a bit 
more involved, therefore, we provide an example.
Consider a technology with $10$ metal layers. Assume that $5$ out of those $10$ 
layers are available for signal routing. Say that out of those $5$ layers, $3$ are used
for horizontal routing and $2$ are used for vertical routing. Assume that the
horizontal layers have a wire pitch of $40$nm, $50$nm, and $60$nm,
and that the vertical layers have a wire pitch of $45$nm and $55$nm.
In this scenario, we would configure the following two functions that are 
explained in the subsequent paragraph:
\begin{equation*}
    f^H_\text{wires $\to$ mm}(x) :=  \frac{x \cdot 10^{-6}}{\frac{1}{40} + \frac{1}{50} + \frac{1}{60}}
    \quad
    f^V_\text{wires $\to$ mm}(x) := \frac{x \cdot 10^{-6}}{\frac{1}{45} + \frac{1}{55}}
\end{equation*}
We start by transforming the wire pitch of each metal layer into the number of wires per nm.
To do this, we compute the reciprocal of the wire pitch. 
Next, we compute the overall number of wires per nm for a given direction which is done by summing up the reciprocal wire pitches of all metal layers for said direction.
Then, we divide the number of wires $x$ by said sum, which gives us the space (in nm) needed to build $x$ wires in a given direction. 
Finally, we multiply the result by $10^{-6}$ to transform it from nm to mm. This computation allows us to represent multiple physical metal layers with
different wire pitches as a single, abstract metal layer.

\begin{table}[h]
\centering
\captionsetup{justification=centering}
\caption{Architectural Parameters Needed as Model Inputs.}
\begin{tabular}{ll}
\hline
\rowcolor{lightgray}
\multicolumn{2}{c}{Parameters describing the chip design}\\
\hline
$N_T$		& Number of tiles\\
$A_E$		& \makecell[l]{Combined area of all endpoints (cores + local
            \\memories) in a tile in gate equivalent (GE)}\\
$R_T$		& Aspect ratio of a tile (height:width)\\
\hline
\rowcolor{lightgray}
\multicolumn{2}{c}{Parameters describing the \gls{noc}}\\
\hline
$F$			& Frequency at which the \gls{noc} is run\\
$B$			& Bandwidth of each router-to-router link\\
\hline
\rowcolor{lightgray}
\multicolumn{2}{c}{Parameters describing the technology node}\\
\hline
$f_\text{GE $\to$ mm$^2$}(x)$			& \makecell[l]{Function; area (in mm$^2$) needed to
                                        \\synthesize $x$ GE of logic.}\\
$f^H_\text{wires $\to$ mm}(x)$			& \makecell[l]{Function; space (in mm) needed to 
								        \\manufacture $x$ parallel, horizontal wires.}\\
$f^V_\text{wires $\to$ mm}(x)$			& \makecell[l]{Function; space (in mm) needed to 
								        \\manufacture $x$ parallel, vertical wires.}\\
$f^L_\text{mm$^2$ $\to$ W}(x)$			& \makecell[l]{Function; approximate power consumption 
								        \\(in W) of $x$ mm$^2$ of logic-dominated area.}\\
$f^W_\text{mm$^2$ $\to$ W}(x)$			& \makecell[l]{Function; approximate power consumption  
								        \\(in W) of $x$ mm$^2$ of wire-dominated area.}\\
$f_\text{mm $\to$ s}(x)$		        & \makecell[l]{Function; time (in s) it takes a signal to
								        \\travel a distance of $x$ mm along a buffered wire.}\\
\hline
\rowcolor{lightgray}
\multicolumn{2}{c}{Parameters describing the on-chip transport protocol}\\
\hline
$f_\text{bw $\to$ wires}(x)$			& \makecell[l]{Function; number of wires needed to build
								        \\a link with a bandwidth of $x$ bits/cycle}\\
$f_{A_R}(m,s,B)$				        & \makecell[l]{Function; area (in GE) of a \gls{noc} router with
								        \\$m$ manager ports, $s$ subordinate ports and bandwidth $B$}\\
\hline
\end{tabular}
\label{tab:model-details-archparam}
\vspace{-1.5em}
\end{table}

\ps{Explain model step-by-step + how to compute outputs}

\subsubsection{Model Details}
\label{sssec:model-custom-details}

We describe the five steps to construct our \gls{noc} model and how to use it to estimate the
\gls{noc}'s area overhead, power consumption and the latency of every router-to-router link.

\begin{figure*}[h]
\vspace{-1.5em}
\centering
\captionsetup{justification=centering}
\begin{subfigure}[t]{0.185 \textwidth}
\centering
\includegraphics[width=1.0\columnwidth]{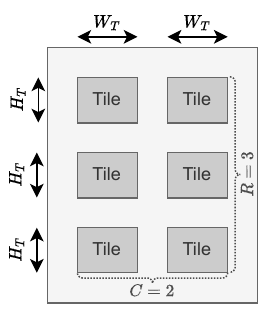}
\caption{Step 1: Tile Area Estimate and Placement in $R \times C$ Grid.}
\label{fig:model-details-step1}
\end{subfigure}
\hspace{-0.6em}
\begin{subfigure}[t]{0.015 \textwidth}
\includegraphics[width=1.0\columnwidth]{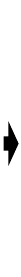}
\end{subfigure}
\hspace{-0.6em}
\begin{subfigure}[t]{0.185 \textwidth}
\centering
\includegraphics[width=1.0\columnwidth]{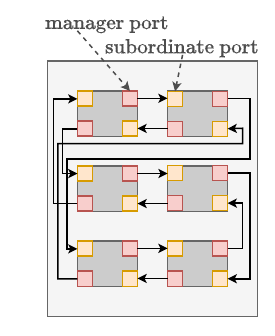}
\caption{Step 2: Global routing in grid of tiles.}
\label{fig:model-details-step2}
\end{subfigure}
\hspace{-0.6em}
\begin{subfigure}[t]{0.015 \textwidth}
\includegraphics[width=1.0\columnwidth]{img/model/arrow.pdf}
\end{subfigure}
\hspace{-0.6em}
\begin{subfigure}[t]{0.185 \textwidth}
\centering
\includegraphics[width=1.0\columnwidth]{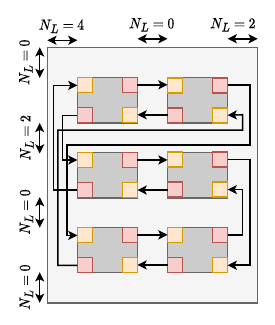}
\caption{Step 3: Estimation of spacing between rows and columns of tiles.}
\label{fig:model-details-step3}
\end{subfigure}
\hspace{-0.6em}
\begin{subfigure}[t]{0.015 \textwidth}
\includegraphics[width=1.0\columnwidth]{img/model/arrow.pdf}
\end{subfigure}
\hspace{-0.6em}
\begin{subfigure}[t]{0.185 \textwidth}
\centering
\includegraphics[width=1.0\columnwidth]{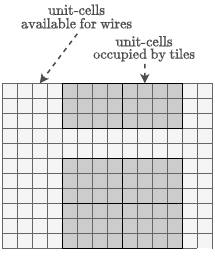}
\caption{Step 4: Discretization of chip into same-sized unit-cells.}
\label{fig:model-details-step4}
\end{subfigure}
\hspace{-0.6em}
\begin{subfigure}[t]{0.015 \textwidth}
\includegraphics[width=1.0\columnwidth]{img/model/arrow.pdf}
\end{subfigure}
\hspace{-0.6em}
\begin{subfigure}[t]{0.185 \textwidth}
\centering
\includegraphics[width=1.0\columnwidth]{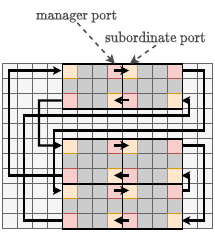}
\caption{Step 5: Detailed routing in the grid of unit-cells.}
\label{fig:model-details-step5}
\end{subfigure}
\caption{The five steps in our model for predictions of area, power consumption, and link 
latencies.}
\label{fig:model-details-steps}
\vspace{-1.5em}
\end{figure*}

\paragraph{Model Construction}
Our model is constructed in five steps:

\begin{itemize}
\item \textbf{Step 1: Tile Area Estimate and Placement in $R \times C$ Grid.}
We start by estimating the area of a tile as $A_T = A_E + A_R$ where $A_R$ is
the area of the tile's local router. The router area is computed as 
$A_R = f_{A_R}(m, s, B)$ where the number of manager-ports $m$ and the number of 
subordinate-ports $s$ both depend on the \gls{noc} topology. The tile height $H_T$ and width $W_T$ 
are computed as follows:
\begin{equation*} \label{eq:model-details-tilssize}
	H_T = \sqrt{R_T \cdot f_\text{GE $\to$ mm$^2$} (A_T)}
	\qquad
	W_T = \sqrt{\frac{f_\text{GE $\to$ mm$^2$} (A_T) }{R_T}}
\end{equation*}
The tiles are then arranged in an $R \times C$ grid (see Figure \ref{fig:model-details-step1}).

\item \textbf{Step 2: Global Routing in Grid of Tiles.}
Since we assume that it is not possible to route links on top of tiles, 
we route them in the space between tiles (see \Cref{fig:model-details-step2}). 
Wire routing in VLSI-design is an NP-complete problem \cite{vlsi-routing-npc}, which is tackled
using heuristic algorithms. Similarly as in real VLSI-design, we use a greedy algorithm to 
perform the global routing. Notice 
that the locations of ports are taken into account in the global routing.

\item \textbf{Step 3: Estimation of Spacing between Rows and Columns of Tiles.}
If there are at most $N_L$ parallel, horizontal links between two rows of tiles, then 
the spacing $S$ between them is estimated as

\begin{equation*} \label{eq:model-step3-tilespacing}
	S = f^H_\text{wires $\to$ mm} \left ( N_L \cdot f_\text{bw $\to$ wires} \left ( B \right ) \right ).
\end{equation*}
To compute the spacing between two columns of tiles with $N_L$ parallel, vertical links 
in between, $f^H_\text{wires $\to$ mm}$ is replaced by $f^V_\text{wires $\to$ mm}$ in the equation
above. Figure \ref{fig:model-details-step3} visualizes this step.
 
\item \textbf{Step 4: Discretization of Chip into same-sized Unit-Cells.}
To facilitate the detailed routing and the computation of area, power and link 
latency estimates, we discretize the chip into same-sized unit-cells. The width $W_{C}$ and 
the height $H_{C}$ of a unit-cell are set such that the unit-cell can accommodate exactly one horizontal and 
one vertical link.
\begin{equation*} \label{eq:model-details-cellsize-h}
	H_{C} = f^H_\text{wires $\to$ mm} \left ( f_\text{bw $\to$ wires} \left ( B \right ) \right )
\end{equation*}
\begin{equation*} \label{eq:model-details-cellsize-w}
	W_{C} = f^V_\text{wires $\to$ mm} \left ( f_\text{bw $\to$ wires} \left ( B \right ) \right )
\end{equation*}
The area of a unit-cell is $A_{C} = H_C \cdot W_C$. Figure \ref{fig:model-details-step4}
displays the chip represented as a grid of unit-cells.

\item \textbf{Step 5: Detailed Routing in Grid of Unit-Cells.}
We use a custom heuristic algorithm to route links in the grid of unit-cells. This 
algorithm tries to reduce the number of collisions (multiple parallel links in the
same unit-cell) and the link lengths (see Figure \ref{fig:model-details-step5}).
\end{itemize}

\paragraph{Area Estimate}
The total area of a chip consisting of $N_\text{cell}$ unit-cells is $A_\text{tot}=N_\text{cell}\cdot A_{C}$.
The area that would be necessary to build the chip without a \gls{noc} which is
\begin{equation*} \label{eq:model-details-arealoc}
	A_\text{noNoC} = F_\text{GE $\to$ mm$^2$} \left ( N_T \cdot A_E \right ).
\end{equation*}
We define the \gls{noc}'s area overhead as the percentage of the total chip area that would be saved
if we would remove the \gls{noc}.
\begin{equation*} \label{eq:model-details-arealoc}
	\text{area\_overhead} = \frac{A_\text{tot} - A_\text{noNoC}}{A_\text{tot}}
\end{equation*}

\paragraph{Power Estimate}
Let $N^L_\text{cell}$, $N^{H}_\text{cell}$, and $N^{V}_\text{cell}$ be the number of unit-cells containing
logic (a tile), a horizontal part of a link and a vertical part of a link 
respectively. Notice that a single unit-cell can contribute to both $N^{H}_\text{cell}$ and 
$N^{V}_\text{cell}$ (if two links are crossing) or it can contribute to none of the three
cell-counts (if it is empty). We estimate the chip's total power consumption as
\begin{equation*} \label{eq:model-details-powertot}
	P_\text{tot} = f^L_\text{mm$^2$ $\to$ W} \left ( N^L_\text{cell} \cdot A_{C} \right ) 
			+ f^W_\text{mm$^2$ $\to$ W} \left ( \left ( N^{H}_\text{cell} + N^{V}_\text{cell} \right )\cdot \frac{A_{C}}{2} \right ).
\end{equation*}
The power consumption of the chip without a \gls{noc} and the \gls{noc}'s power consumption are computed 
as follows:
\begin{equation*} \label{eq:model-details-powernocnoc}
	P_\text{noNoC} = f^L_\text{mm$^2$ $\to$ W} \left ( f_\text{GE $\to$ mm$^2$} \left ( N_T \cdot A_E \right ) \right ) 
\end{equation*}
\begin{equation*} \label{eq:model-details-powernoc}
	P_\text{NoC} = P_\text{tot} - P_\text{noNoC}
\end{equation*}

\paragraph{Link Latency Estimate}
We estimate the latency $L$ (in clock cycles) of a link that crosses 
$N^H_\text{cell}$ unit-cells vertically and $N^V_\text{cell}$ unit-cells horizontally as follows:

\begin{equation*} \label{eq:model-details-conlatency}
	L = f_\text{mm $\to$ s} \left ( N^H_\text{cell} \cdot W_C + N^V_\text{cell} \cdot H_C \right ) \cdot F.
\end{equation*}

\ps{Assess toolchain accuracy}

\subsection{Toolchain Evaluation}
\label{ssec:model-eval}
To assess the accuracy of our prediction toolchain, we use it to predict the performance
and cost of the open-source MemPool \cite{mempool} architecture. Besides having a complex low-latency
interconnect between its $256$ cores and $1024$ memory banks, MemPool also has a long implementation runtime. This makes it a prime target for our model, which can give information about the performance and cost of a certain topology without requiring the long iterative place-and-route flow.  Table \ref{tab:model-accuracy} shows
the cost and performance numbers of MemPool, the predictions of our toolchain, and its 
prediction error. The area and power predictions are accurate for a fast high-level 
model. Our model over-estimates the latency of MemPool because MemPool is highly
latency-optimized, which breaks our assumptions that each link and each router has
a minimum latency of one cycle. To take this into account, we need to deduct $4$ cycles from
our model estimate ($1$ cycle to inject the flit and $1$ cycle for each of the three routers
that a given flit traverses). This would result in a latency estimate of $6$ cycles, which is 
only off by $20\%$. For most designs, our assumptions on the router latency 
hold, and such corrections are not necessary. 

\begin{table}[h]
\centering
\captionsetup{justification=centering}
\caption{Cost and Performance results and predictions of the MemPool Architecture \cite{mempool}.}
\begin{tabular}{lccc}
\hline
\rowcolor{lightgray}
Metric 					& Correct Value & Prediction 		& Prediction Error \\
\hline
Area					& $21.16$ mm$^2$	& $24.26$ mm$^2$	& $15$\%			\\
Power 					& $1.55$ W		    & $1.447$ W			& $7$\%			\\
Latency 				& $5$ cycles	    & $10$ cycles		& $100$\%			\\
Throughput				& $38$\%		    & $25$\%			& $34$\%			\\
\hline
\end{tabular}
\label{tab:model-accuracy}
\vspace{-1.5em}
\end{table}

\section{Application and Evaluation}
\label{sec:evaluation}

\picom{ToDo: Discuss that the evaluation supports our design principles}

\ps{Section intro}

We show how to use our prediction toolchain and design principles to
customize the sparse Hamming graph topology.
To assess the quality of our customization strategy, we compare customized
sparse Hamming graph topologies to a series of established NoC topologies.

\ps{Explain how toolchain and design principles are used to customize the sparse Hamming graph topology}

\paragraph{\gls{noc} Topology Customization Strategy}

\begin{itemize}
    \item Step 1: Start with the simplest sparse Hamming graph topology, which is a mesh
    ($S_R = \{\}$, $S_C = \{\}$). 
    \item Step 2: Use our prediction toolchain to estimate the performance and cost of the current topology 
    if applied to the target architecture with its unique architectural parameters. 
    \item Step 3: Compare the cost and performance estimates from step 2 to the design 
    goals to identify insufficiencies of the current topology (e.g., inadequate throughput). 
    \item Step 4: Follow our design principles to change the parameters $S_R$ and $S_C$ such that the
    insufficiencies identified in step 3 are eliminated (e.g., by adding additional links to reduce the network 
    diameter, and hence improve the throughput).
    \item Step 5: Go back to step 2) and repeat the process until you are satisfied with the performance
    and cost of the customized topology.
\end{itemize}

\ps{Introduce architectural parameters for evaluation}

\paragraph{Target Architectures for Evaluation}
Assume that we are tasked with customizing a \gls{noc} topology for an architecture similar to 
Knights Corner (KNC). We select this architecture because it is implemented in a 
$22$ nm technology node for which we know the necessary architectural parameters.
This means that we want to connect $64$ tiles (KNC has $62$) with an area of 
approximately $35$ MGE each using a \gls{noc} with $512$ bits/cycle per-link bandwidth running 
at $1.2$ GHz. We assume that the AXI \cite{ambaaxi} transport protocol and 
the AXI-NoC-components from Kurth et al. \cite{axinoc} are used. Furthermore, we use 
input-queued routers with $8$ virtual channels and $32$-flit buffers.\picom{I also tried $4$ VCs 
as suggested by Tim but with only $4$ VCs many topologies deadlock at 10\% load...}
Let this be scenario a). 
We analyze three additional scenarios for scaling up the 
number of cores: Scenario b) increase the number of cores per tile by $2 \times$,
scenario c) increase the number of tiles by $2 \times$, and scenario d) increase
the number of cores per tile and the number of tiles by $2 \times$. 
Assume that for all 4 scenarios, our design goal is to maximize the 
global bandwidth (priority 1) and minimize the latency (priority 2) without exceeding a 
\gls{noc} area overhead of $40\%$. 

\ps{Discuss Figure \ref{fig:evaluation-results}}

\paragraph{Evaluation Results}
We apply our customization strategy to all four scenarios explained in the previous paragraph.
Figure \ref{fig:evaluation-results} shows a cost and performance comparison of our customized 
sparse Hamming graph topologies against a series of established topologies. The topologies
are compared in terms of four metrics: area overhead, power consumption, saturation
throughput, and zero-load latency. As a consequence, a topology is usually not strictly better 
than another topology, instead, each topology reaches s certain trade-off between those
four metrics. In the previous paragraph, we clearly define our design goal:
Maximize the throughput and minimize the latency without exceeding an area overhead of $40\%$.
The sparse Hamming graph topology allows us to increase the 
performance by adding additional links until an area overhead of  $40\%$ is reached. As a result,
our customized sparse Hamming graph topologies deliver the highest global throughput
among all topologies with an area overhead of at most $40\%$ while providing the 
second lowest latency among all topologies considered. We conclude that while there
is no single topology that is superior with respect to all four metrics, the sparse Hamming 
graph is the only topology where the cost-performance trade-off can be adjusted based on the design goals.

\begin{figure*}[t]
\centering
\captionsetup{justification=centering}
\begin{subfigure}{0.98 \textwidth}
\centering
\includegraphics[width=1.0\columnwidth]{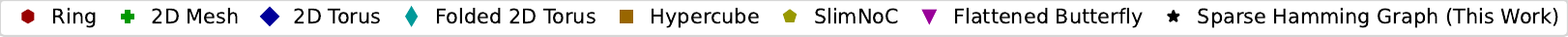}
\end{subfigure}
\begin{subfigure}{0.48 \textwidth}
\centering
\includegraphics[width=1.0\columnwidth]{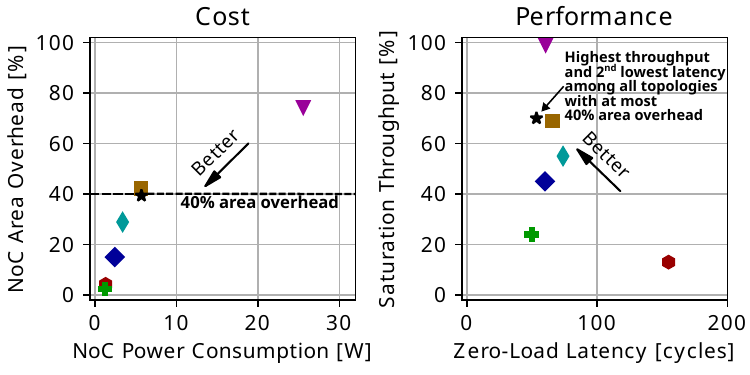}
\vspace{-1.5em}
\caption{64 tiles with 35MGE and 1 core each. Parameters for sparse Hamming graph: $S_R = \{4\}$, $S_C = \{2,5\}$.}
\label{fig:intro-comparison-setting1}
\end{subfigure}
\begin{subfigure}{0.48 \textwidth}
\centering
\includegraphics[width=1.0\columnwidth]{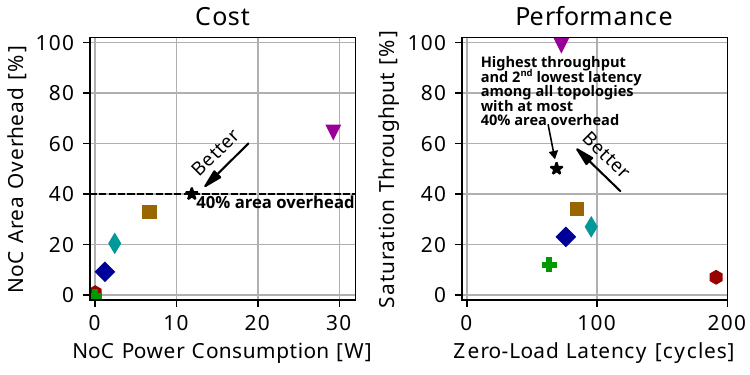}
\vspace{-1.5em}
\caption{64 tiles with 70MGE and 2 cores each. Parameters for sparse Hamming graph: $S_R = \{2,4\}$, $S_C = \{2,4\}$.}
\label{fig:intro-comparison-setting2}
\end{subfigure}
\begin{subfigure}{0.48 \textwidth}
\centering
\includegraphics[width=1.0\columnwidth]{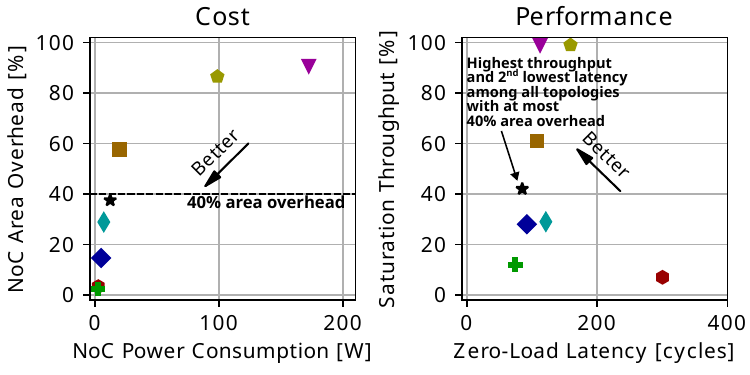}
\vspace{-1.5em}
\caption{128 tiles with 35MGE and 1 core each. Parameters for sparse Hamming graph: $S_R = \{3\}$, $S_C = \{2,5\}$.}
\label{fig:intro-comparison-setting1-large}
\end{subfigure}
\begin{subfigure}{0.48 \textwidth}
\centering
\includegraphics[width=1.0\columnwidth]{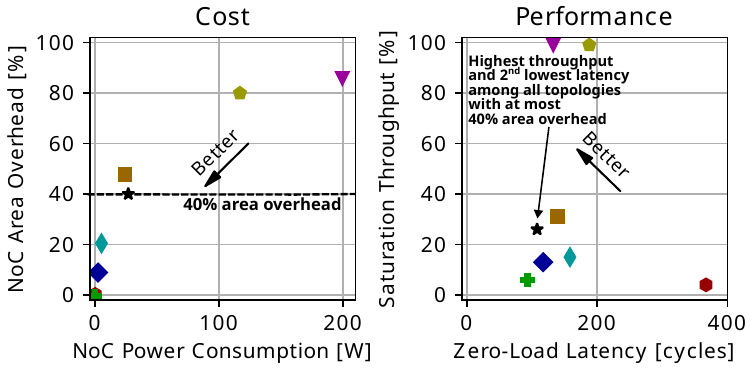}
\vspace{-1.5em}
\caption{128 tiles with 70MGE and 2 cores each. Parameters for sparse Hamming graph: $S_R = \{2,4\}$, $S_C = \{2,4\}$.}
\label{fig:intro-comparison-setting2-large}
\end{subfigure}
\caption{Comparison of various topologies for four different scenarios. 
The comparison is performed using our prediction toolchain with a random uniform traffic 
pattern and a routing algorithm that minimizes the number of router-to-router hops.
SlimNoC is only applicable for scenarios c) and d) because it requires the number of tiles $N_T$ to be 
$N_T=2p^2$ for a prime power $p$.}
\label{fig:evaluation-results}
\vspace{-1.5em}
\end{figure*}

\section{Related Work}
\label{rel-work}

\ps{Related work: Topologies}

State-of-the-art manycore architectures \cite{knl, marvellthunderx3} often use ring or mesh topologies with a low router radix and a high network diameter.
These topologies come at a low cost, but their performance steeply deteriorates when scaling them up from dozens to hundreds or even thousands of cores.
New topologies such as Flattened Butterfly \cite{flattened-butterfly-noc} or SlimNoC \cite{slimnoc} alleviate this performance degradation 
by reducing the network diameter, albeit with a higher router radix and cost when scaled. 
Topologies such as 2D torus, folded 2D torus \cite{folded-torus}, or hypercube \cite{hypercube} try to fill the gap between these two groups of topologies, 
but they can not be customized to adjust their cost-performance trade-off.
Ruche networks \cite{ruche-nw} are mesh-based networks with length-adjustable skip-links, however, the number of configurations is quite limited.
Sparse Hamming graphs are a superset of Ruche networks and they provide significantly more configurations than Ruche networks.
As a consequence, sparse Hamming graphs offer a more fine-grained adjustment of the cost-performance trade-off. 

\ps{Related work: Cost and Performance Prediction}

There are several high-level models \cite{booksim, topaz, garnet} performing cycle-level 
simulations for performance predictions. High-level area and power predictions are also 
possible, e.g., by using Orion 2.0 \cite{orion2}.
However, these schemes do not consider implementation details such as link routing,
hence, they suffer from limited accuracy. 
There are also low-level models \cite{opensmart, nocgen, connect} that generate \gls{rtl} 
descriptions of the \gls{noc}. These \gls{rtl} descriptions are used for accurate performance and cost predictions. 
Unfortunately, \gls{rtl} modeling is time-consuming, requiring many person-months to explore a few topologies. 
This amount of effort renders low-level models impractical for 
the customization of \gls{noc} topologies where one wants to quickly explore a large design space. 
PyOCN \cite{pyocn} provides a unified framework for both high-level and low-level predictions,
however, even they cannot combine the low execution time of high-level models with the high
accuracy of low-level models. Our prediction toolchain works at the speed of high-level models while 
estimating low-level details by performing approximate floorplanning and link routing.
\section{Conclusion}
\label{sec:conclusion}

\ps{Motivation for customizable NoC topologies}

The rising number of compute cores per chip and the fact that more and more applications
are becoming bound by data movement makes scalable \gls{noc} topologies more
important than ever. As high performance and low cost can not be unified,
we need topologies with a customizable cost-performance trade-off.

\ps{Our contributions}

We improve the \gls{noc} topology customization process by making three major contributions: 
1) a set of \gls{noc} topology design principles that reveal how to influence the NoC's cost and performance,
2) the customizable sparse Hamming graph topology with an adjustable cost-performance trade-off, and
3) a fast and easy-to-use toolchain for predicting the \gls{noc}'s performance and cost featuring
our custom model for area, power, and link latency estimates.

\ps{Why are our contributions valuable?}

We show that by applying all three contributions, one is able to achieve 
desired cost-performance trade-offs while outperforming established topologies in terms 
of cost, performance, or both.

\ifblind
\else
\section*{Acknowledgements}
\label{sec:ack}

This work was supported by the ETH Future Computing Laboratory (EFCL), financed by a donation from Huawei Technologies.
It also received funding from the European Research Council
\raisebox{-0.25em}{\includegraphics[height=1em]{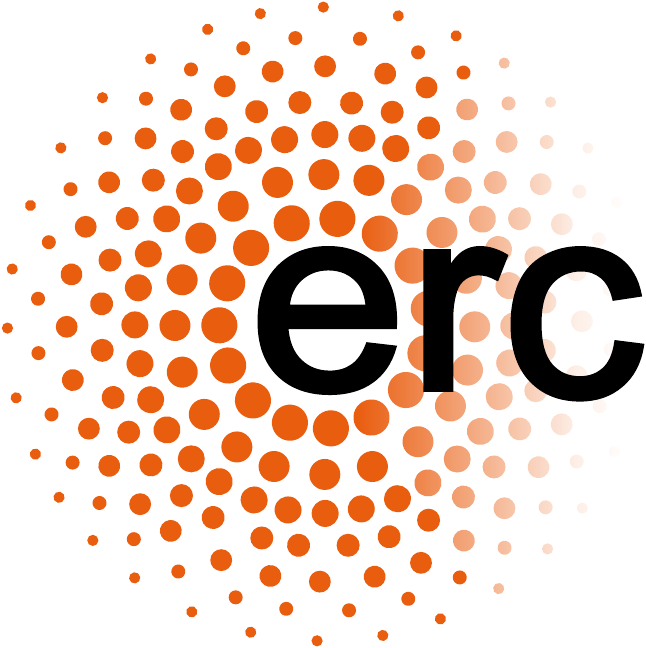}} (Project PSAP,
No.~101002047) and from the European Union's HE research 
and innovation programme under the grant agreement No.~101070141 (Project GLACIATION).
We thank Timo Schneider for help with computing infrastructure at SPCL.
We thank Florian Zaruba for help in the initial stages of the project.
\fi

\bibliographystyle{IEEEtran}
\bibliography{bibliography}

\end{document}